\documentclass[pra,superscriptaddress,showpacs]{revtex4}
\usepackage{amsfonts}
\usepackage{amsmath}
\usepackage{graphicx}

\begin{document}

\title{Variational properties of a pumped dynamical system}
\author{Erika Andersson}
\affiliation{SUPA, School of Engineering and Physical Sciences,
 Heriot-Watt University, Edinburgh EH14 4AS, United Kingdom}
\author{Stig Stenholm}
\affiliation{Physics Department, Royal Institute of Technology,
KTH, Stockholm, Sweden}
\affiliation{Laboratory of Computational Engineering,
HUT, Espoo, Finland}

\begin{abstract}
We have earlier constructed a generalized entropy concept to show the direction 
of time in an evolution following from a Markov generator. In such a dynamical 
system, the entity found changes in a monotonic way starting from any initial 
state of the system. In this paper, we generalize the treatment to the case when 
population is pumped into the system from levels not explicitly considered. These 
populations then pass through the coupled levels and exit by decay to levels 
outside the system. We derive the form of the equation of motion and relate it to 
our earlier treatments. It turns out that the formalism can be generalized to the 
new situation. Its physically relevant features are demonstrated, and the 
behaviour obtained is illustrated by numerical treatment of the standard 
two-level system with pumping and relaxation included.
\end{abstract}

\maketitle

\section{\protect\bigskip Introduction}

All irreversible evolution models should carry some signature of the direction 
of time. In the theory of dynamical systems \cite{Q5} such an entity is called a 
Lyapunov function. This defines uniquely the forward direction of time. In a 
physical context, the function may be considered to be a generalization of the 
entropy concept. In the case of approach to thermal equilibrium, the description 
should consistently reproduce what we take to be known from thermodynamics. 
The procedure for finding this description will, of course, depend on the model 
we offer for the time evolution. The general concept of the direction 
of physical time is discussed rather comprehensively in \cite{Q5a}.

The modern theory of kinetic equations is compiled in the text \cite{Q6}. The 
quantum counterparts are presented in \cite{Q7}. Recent progress in quantum 
dynamics has created a separate research field named \textit{open systems 
dynamics} \cite{Q8}; its more mathematical aspects are presented in 
\cite{Q9}.

The contemporary focus on information measures has led to the introduction 
of a multitude of generalized entropies \cite{Q12} and \cite{Q13}. The various 
expressions have different properties and different uses. In our earlier work, 
we have attempted to describe the evolution of a general dynamical system 
by constructing a state functional displaying the direction of time by monotonically 
changing. This may then be regarded as a generalized entropy or, alternatively, 
as a Lyapunov function.

In general, the time evolution equation of a physical system is of the first
order and generated by a linear operator on the object defining the state of
the system. For convenience, the generator is here taken to be time
independent, which restricts the applicability of our approach. The time
step is taken to depend only on the state at the present time. This is
usually called a \textit{Markovian evolution} in physics. Within these
restrictions we have been able to develop a rather detailed theory for the
general case.

In \cite{Q14} the solution was presented for a restricted class of evolution
equations. In the case of a driven two-level system with decay, the solution
was presented in \cite{Q15}. As the cases discussed that far assumed no
degeneracy, this was treated in \cite{Q16}. The case of a thermal reservoir
was introduced in \cite{Q17}.

The method introduced by us displays a monotonically evolving quantity. This
serves to define a forward direction of time. In our paper \cite{Q18}, we
derive the proper time-inversed behavior of \ irreversible time evolution.
The related minimum property of entropy production is discussed in \cite{Q18a}.

In this paper, we want to apply the theory to a case not usually addressed
by the formal evolution theory. We consider the case where the physics takes
place in a state space pumped by externally induced means. The population in
the space considered is thus increasing with time, but this is counteracted
by decay channels out of it. This system is not of the canonical Lindblad
form \cite{Q30}, but it retains the physical interpretation by letting the
density matrix elements denote the density of active particles instead of
the customary probabilities. In laser physics and nonlinear spectroscopy
this system has frequently been taken to model the physical situation
investigated \cite{Q30a}. As most experimental laser physics is still described by equations of this
class, we regard it to be important to relate such equations to the more
recent approach developed by us.

In Sec. 2, we introduce the concept of a pumped system and its quantum
mechanical description \ by a linear master equation. In Sec. 3 we discuss
the physical situation as it differs from a genuine Lindblad case. Finally
we apply the theory introduced earlier to this case, and extract its
characterization of the direction of time. The theory is then, in Sec. 4,
illustrated by \ applying it to the case of a pumped decaying two-level
system. This model has played a central role in both laser spectroscopy and
quantum optics \cite{Q31}. The formalism, however, also applies to Markovian
rate equations \cite{Q33}. Finally the conclusions from the treatment are
discussed in Sec. 5.

\section{The pumped system}

\subsection{The time evolution equation}

The terminology of quantal state evolution assigns the term \textit{open
system} to any quantum system in contact with an environment.
Ordinarily the system is
probabilistically closed, i.e. we assume that the density matrix of the
system retains the probability interpretation and is normalized.
Consequently all probability exchange between the states, reversible as well
as dissipative, occurs between levels in the system. The general memory-less
expression for such evolution is given by the Lindblad form of the
generating equation.

There is, however, a class of time evolution equations that do not conform
to this model. There \ is a flow of probability through the system; the
states are pumped by some technical method and the influx of probability is
compensated by states decaying to unobserved lower levels. Such models are
termed \textit{pumped systems }here\textit{. }Naturally the state space
could be enlargened so that all pumping and decay mechanisms are included in
the dynamics of the evolution. It is, of course, always possible to consider
a space large enough to eliminate the need for a phenomenological master
equation, but it is not always an expedient or even a practical possibility.

The time evolution equation will now hold several different pieces:

\begin{enumerate}
\item The decays inside and out of the system are described by a relaxation
operator  
$\mathcal{R}$, which we assume to be linear and memoryless. The time 
derivative of the density matrix in the absence of pumping and with $H=0$, 
where $H$ is the Hamilton operator, would be
\begin{equation}
\partial _{t}\rho =\mathcal{R}\rho .   \label{e0}
\end{equation}

\item The pumping is described by a random process which at time $t_{0}$
introduces a system in the state%
\begin{equation}
\mid \psi \rangle =\sum_{n}c_{n}\left( 0\right) \mid \varphi _{n}\rangle ,
\label{e1}
\end{equation}%
where $\left\{ \mid \varphi _{n}\rangle \right\} $ is an arbitrary basis on
the state space of the system under consideration. At a later time this will, in 
the absence of relaxation,
evolve to the state%
\begin{equation}
\mid \psi \left( t\right) \rangle =\sum_{n}c_{n}\left( t-t_{0}\right) \mid
\varphi _{n}\rangle .  \label{e2}
\end{equation}%
We now form an ensemble accumulated by all systems introduced at times $%
t_{0}\leq t.$ This is described by the state%
\begin{equation}
\tilde{\rho}_{nm}(t)=\int\limits_{-\infty }^{t}c_{n}\left( t-t_{0}\right)
c_{m}^{\ast }\left( t-t_{0}\right) P_{\psi }(t_{0})dt_{0},  \label{e3}
\end{equation}%
where $P_{\psi }(t_{0})$ is the probability of introducing a system in state 
$\psi $ into the ensemble at time $t_{0}.$ This is assumed to be normalized
over the states introduced:%
\begin{equation}
\sum_{\psi }P_{\psi }(t)=1.  \label{e3a}
\end{equation}%
The contribution from (\ref{e3}) to the time evolution is given by two terms,%
\begin{equation}
\begin{array}{lll}
\partial _{t}\tilde{\rho}_{nm}(t) & = & c_{n}\left( 0\right) c_{m}^{\ast
}\left( 0\right) P_{\psi }(t)+\int\limits_{-\infty }^{t}\partial _{t}\left[
c_{n}\left( t-t_{0}\right) c_{m}^{\ast }\left( t-t_{0}\right) \right]
P_{\psi }(t_{0})dt_{0}.%
\end{array}
\label{e4}
\end{equation}%
We now introduce the density operator averaged over the pumped states 
in (\ref {e1}) by setting%
\begin{equation}
\sum_{\psi }\tilde{\rho}_{nm}(t)\rightarrow \rho _{nm}(t).  \label{e3b}
\end{equation}%
We may then write the averaged time evolution as%
\begin{equation}
\partial _{t}\rho _{nm}(t)=\Lambda _{nm}(t)+C_{nm}(t),  \label{e4a}
\end{equation}%
where the inhomogeneous term is given by%
\begin{equation}
\Lambda _{nm}(t)=\sum_{\psi }c_{n}\left( 0\right) c_{m}^{\ast }\left(
0\right) P_{\psi }(t).  \label{e4b}
\end{equation}%
Because the Hamiltonian is taken to be the same for all systems pumped into
the ensemble, we let the state coefficients satisfy the ordinary Schr\"{o}%
dinger equation in the form%
\begin{equation}
\partial _{t}c_{n}\left( t-t_{0}\right) =-i\sum_{k}H_{nk}c_{k}\left(
t-t_{0}\right) ,  \label{e5}
\end{equation}%
where $H$ is the generator of the unitary time evolution inside the system.
Using this in (\ref{e4}) we obtain%
\begin{equation}
C_{nm}(t)=-i\left( \left[ H,\rho (t)\right] \right) _{nm},  \label{e6}
\end{equation}%
where we have introduced the ensemble density matrix from (\ref{e3b}).
\end{enumerate}

Now, collecting terms from (\ref{e0}) and (\ref{e4a}), and using (\ref{e6}), we find the
time evolution equation for the pumped ensemble to be
\begin{equation}
\partial _{t}\rho =\Lambda (t)-i\left[ H,\rho (t)\right] +\mathcal{R}\rho
(t).  \label{e7}
\end{equation}%
This is the most general linear and memoryless evolution equation we
encounter. The Lindblad form falls under this more general class if we set 
$\Lambda (t)=0.$ If this is not the case, then the relaxation term $%
\mathcal{R\rho }$ cannot be of the Lindblad form, which on its own would
ensure conservation of the probability. As the term $\Lambda $ does not do
this, its influence must be compensated by $\mathcal{R}\rho .$

Not being of the Lindblad form, Eq. (12) does not manifestly generate
completly positive time evolution. However, properly used, it does not
violate the physical interpretation of the density matrix. The pumping in
Eq. (9) is of the form of a density matrix and thus cannot cause any
troubles. The damping operator (1) has to be introduced such that no
abnormal results emerge. This implies inequalities between the decay
constants, but these are no different from the corresponding relations
encountered in Markovian rate processes, Ref. [18]. The Hamiltonian
evolution, naturally, induces only physically acceptable changes.

\subsection{Properties of pumped evolution}

The equation (\ref{e7}) may be written in a notation generalizing the
Lindblad case as%
\begin{equation}
\partial _{t}\rho =\Lambda +\mathcal{L}\rho .  \label{e8}
\end{equation}%
In agreement with the concept of pumping a state, it follows that%
\begin{equation}
\Lambda _{nn}=\sum_{\psi }c_{n}\left( 0\right) c_{n}^{\ast }\left( 0\right)
P_{\psi }(t)\geq 0;  \label{e9}
\end{equation}%
this term can only add to the population of the state. The pumping mechanism
is usually incoherent, and $\Lambda $ is not expected to have any
non-diagonal contributions, because of the ensemble average.

Conservation of probability requires that the steady state, if it exists, satifies 
\begin{equation}
Tr\left( \mathcal{R}\rho _{0}\right) =-Tr\Lambda .  \label{e10}
\end{equation}%
The operator $\mathcal{R}$ is thus essentially a negative operator.

If no pumping occurs, $\Lambda =0,$ there are two possibilities:

\begin{itemize}
\item After an infinite time, all population decays out of the system and
the only steady state is the one with all populations vanishing, $\rho
_{0}(\infty )=0.$ In this situation, the relation%
\begin{equation}
\mathcal{L}\rho =0  \label{e11}
\end{equation}%
has got no nonvanishing solutions; all eigenvalues of $\mathcal{L}$ are
nonzero.

\item There is a part of $\mathcal{L}$ which allows for a steady state with
nonvanishing density matrix%
\begin{equation}
\mathcal{L}_{0}\rho ^{0}=0;\;\rho ^{0}\neq 0.  \label{e12}
\end{equation}%
This concerns a part of the state space which has no decay channel out of
it, and population ending up here will remain trapped. Thus no part of the
pumped population can go into this subspace, because then probability would
accumulate here without limit. It thus describes a part of the state space
which is totally decoupled from the rest of the system: No coherent transfer
nor any irreversible decay can couple this to the rest. It is consequently a
system which can be treated by the ordinary theory; all its population
derives from its initial state and no pumping will affect it. In the
subsequent discussion we omit this possibility.
\end{itemize}

The steady state solution of (\ref{e8}) becomes%
\begin{equation}
\rho _{0}=-\mathcal{L}^{-1}\Lambda . \label{e13}
\end{equation}%
According to what is said above, the operator $\mathcal{L}$ has got no zero
eigenvalues, and the expression (\ref{e13}) will exist in all physically
acceptable situations.

The ensemble density matrix elements $\rho _{nn}\geq 0$ all denote the
population on level $n$ but they no longer necessarily relate to any
probability; in particular it need not hold that 
\begin{equation}
Tr\rho _{0}=Tr\rho (t).  \label{e14}
\end{equation}%
Note that the special initial condition, $\rho (t_{0})=0$, is perfectly
acceptable and may still imply $\rho _{0}=\rho (\infty )\neq 0.$ The
off-diagonal elements of $\rho $ retain their role as determining the
multipole moments of the system.

With this notation we may write (\ref{e8}) as%
\begin{equation}
\partial _{t}\rho =\mathcal{L}\left( \rho -\rho _{0}\right) ;  \label{e15}
\end{equation}%
with the notation%
\begin{equation}
\delta \rho \equiv \rho -\rho _{0} . \label{e16}
\end{equation}%
This allows us to write%
\begin{equation}
\partial _{t}\delta \rho =\mathcal{L\delta }\rho ,  \label{e17}
\end{equation}%
which relates the present case to our earlier theory.

\section{Physics of time evolution}

\subsection{\protect\bigskip Formal properties}

The objects $\rho $ representing the state of the system obey the general
dynamic time evolution equation of the form 
\begin{equation}
\partial _{t}\rho = \partial_t \delta\rho = \Lambda +\mathcal{L}\rho =
\mathcal{L\delta }\rho .
\label{a1}
\end{equation}%
They can be taken to belong to a linear manifold of elements denoted by $%
\mid \rho \rangle \rangle .$ This space can be equipped with a natural inner
product by writing 
\begin{equation}
\langle \langle \rho _{1}\mid \rho _{2}\rangle \rangle \equiv Tr(\rho
_{1}^{\dagger }\rho _{2}).  \label{a3}
\end{equation}

The time evolution operator is not assumed to be Hermitian with respect to
the inner product defined. Consequently we need to introduce right
eigenvectors 
\begin{equation}
\mathcal{L}\mid x_{\nu }\rangle \rangle =\lambda _{\nu }\mid x_{\nu }\rangle
\rangle  \label{a4}
\end{equation}%
and left eigenvectors 
\begin{equation}
\mathcal{L}^{\dagger }\mid y_{\nu }\rangle \rangle =\lambda _{\nu }^{\ast
}\mid y_{\nu }\rangle \rangle .  \label{a5}
\end{equation}%
From these relations follows that 
\begin{equation}
\begin{array}{ccc}
\langle \langle y_{\mu }\mid \mathcal{L}\mid x_{\nu }\rangle \rangle & = & 
\lambda _{\nu }\langle \langle y_{\mu }\mid x_{\nu }\rangle \rangle \\ 
&  &  \\ 
\langle \langle \mathcal{L}^{\dagger }y_{\mu }\mid x_{\nu }\rangle \rangle & 
= & \lambda _{\mu }\langle \langle y_{\mu }\mid x_{\nu }\rangle \rangle .%
\end{array}
\label{a5a}
\end{equation}%
Thus if $\lambda _{\nu }\neq \lambda _{\mu }$ the states are mutually
orthogonal and may be normalized against each other%
\begin{equation}
\langle \langle y_{\nu }\mid x_{\mu }\rangle \rangle =\delta _{\nu \mu }.
\label{a5b}
\end{equation}%
We assume in the following that both the right and left eigenstates form
complete basis sets. This is, in particular, true if their corresponding
eigenvalues are nondegenerate, but in general they are complex.

With these definitions, we can present spectral representations for the
operators 
\begin{equation}
\begin{array}{l}
\mathcal{L}=\sum_{\nu }\mid x_{\nu }\rangle \rangle \lambda _{\nu
}\langle \langle y_{\nu }\mid \\ 
\\ 
\mathcal{L}^{\dagger }=\sum_{\nu }\mid y_{\nu }\rangle \rangle
\lambda _{\nu }^{\ast }\langle \langle x_{\nu }\mid .%
\end{array}
\label{a7}
\end{equation}%
When both sets of eigenstates are complete, the identity operator can be
written as 
\begin{equation}
\mathfrak{I}=\sum_{\nu }\mid x_{\nu }\rangle \rangle \langle \langle
y_{\nu }\mid =\sum_{\nu }\mid y_{\nu }\rangle \rangle \langle
\langle x_{\nu }\mid .  \label{a8}
\end{equation}

The inner product between states belonging to the right eigenvalues have no
simple relations. In order to obtain a simple situation, we introduce the
mapping 
\begin{equation}
\Omega :\{\mid x_{\nu }\rangle \rangle \}\Rightarrow \{\mid y_{\nu }\rangle
\rangle \}  \label{a8a}
\end{equation}%
by setting 
\begin{equation}
\Omega =\sum_{\nu }\mid y_{\nu }\rangle \rangle \langle \langle
y_{\nu }\mid .  \label{a9}
\end{equation}%
The inverse mapping is seen to be 
\begin{equation}
\Omega ^{-1}=\sum_{\nu }\mid x_{\nu }\rangle \rangle \langle \langle
x_{\nu }\mid .  \label{a10}
\end{equation}%
These are positive Hermitian operators, and it is possible to define a new
metric based on the bilinear form defined as 
\begin{equation}
M_{\Omega }[\rho _{1},\rho _{2}]\equiv \langle \langle \rho _{1}\mid \Omega
\mid \rho _{2}\rangle \rangle \equiv Tr(\rho _{1}^{\dagger }\Omega \rho
_{2}).  \label{a11}
\end{equation}%
This has got all the properties of an inner product, and it thus defines a
topology in the space of all quantum states. Within this product we have 
\begin{equation}
M_{\Omega }[\mid x_{\nu }\rangle \rangle ,\mid x_{\mu }\rangle \rangle
]=\langle \langle x_{\nu }\mid y_{\mu }\rangle \rangle =\delta _{\nu \mu }.
\label{a12}
\end{equation}%
A similar construction is possible on the states $\{\mid y_{\nu }\rangle
\rangle \}$ by the use of $\Omega ^{-1}.$The operator $\Omega $ may be
considered as a \textit{metric operator }on the manifold of physical states.

By direct calculation we find that 
\begin{equation}
\Omega \mathcal{L}\Omega ^{-1}=\sum_{\nu }\mid y_{\nu }\rangle
\rangle \lambda _{\nu }\langle \langle x_{\nu }\mid =\mathcal{L}^{\ast
\dagger }.  \label{a13}
\end{equation}%
Here as in the following, we assume that complex conjugation denoted by *
affects only c-numbers, not states.

From (\ref{a13}), we derive the relations 
\begin{equation}
\begin{array}{lll}
\Omega \mathcal{L}=\mathcal{L}^{\ast \dagger }\Omega & ; & \mathcal{L\,}%
\Omega ^{-1}=\mathcal{\,}\Omega ^{-1}\mathcal{L}^{\ast \dagger } \\ 
&  &  \\ 
\Omega \mathcal{L}^{\ast }=\mathcal{L}^{\dagger }\Omega , & ; & \mathcal{L}%
^{\ast }\Omega ^{-1}=\Omega ^{-1}\mathcal{L}^{\dagger },%
\end{array}
\label{a14}
\end{equation}%
forming the basis for the considerations below.

If $\mathcal{L}$ has real eigenvalues, then $\mathcal{L}^{\ast }=\mathcal{L}$ 
and the operator is Hermitian with respect to the new inner product 
$M_{\Omega }$ defined above.
If the eigenvalues occur in complex conjugate pairs, i.e. for each $\mid
x_{\mu }\rangle \rangle $ there exists an $\mid x_{\mu }^{\ast }\rangle
\rangle $ such that 
\begin{equation}
\mathcal{L}\mid x_{\mu }^{\ast }\rangle \rangle =\lambda _{\mu }^{\ast }\mid
x_{\mu }^{\ast }\rangle \rangle ,  \label{a14a}
\end{equation}%
we find from 
\begin{equation}
\lambda _{\mu }\,\langle \langle y_{\mu }\mid x_{\mu }^{\ast }\rangle
\rangle =\langle \langle y_{\mu }\mid \mathcal{L}\mid x_{\mu }^{\ast
}\rangle \rangle =\lambda _{\mu }^{\ast }\,\langle \langle y_{\mu }\mid
x_{\mu }^{\ast }\rangle \rangle ,  \label{a14b}
\end{equation}%
that 
\begin{equation}
\langle \langle y_{\mu }\mid x_{\mu }^{\ast }\rangle \rangle \equiv \langle
\langle x_{\mu }\mid \Omega \mid x_{\mu }^{\ast }\rangle \rangle =0;
\label{a14c}
\end{equation}%
these states are thus orthogonal in the $M_{\Omega }$ metric. We expect the
eigenvalues of physical evolution operators to correspond to damped
oscillations at frequencies given by the energy differences in the system.
We call this case \textit{physical evolution}.

The time dependent state that is a solution of the evolution equation (\ref%
{a1}) is given by 
\begin{equation}
\mid \delta\rho (t)\rangle \rangle =\sum_{\nu }r_{\nu }\exp \left( \lambda
_{\nu }t\right) \mid x_{\nu }\rangle \rangle ,  \label{a15}
\end{equation}%
where 
\begin{equation}
r_{\nu }=\langle \langle y_{\nu }\mid \delta\rho (0)\rangle \rangle .  \label{a16}
\end{equation}

From the physical meaning of the time evolution operator in a pumped system,
we derive the following conclusions:

\begin{itemize}
\item There exists no steady state $\mid x_{0}\rangle \rangle $ with
eigenvalue $\lambda _{0}=0.$ Asymptotically, the time evolution will lead to 
$\rho (\infty )=\rho _{0}.$

\item Any solution must approach this in a smooth manner, thus we require $%
{\rm Re}~\lambda _{\nu }<0$ for all $\nu .$
\end{itemize}

\subsection{The direction of time}

As the system undergoes irreversible time evolution, it singles out one
direction of time defining the forward progress. In classical dynamics, we
can achieve this by introducing a monotonically changing variational
function; in system theory this is called a Lyapunov function.

We find easily that the definition 
\begin{equation}
M_{\Omega }(\delta\rho ,\delta\rho )=Tr\left( \delta \rho ^{\dagger }\Omega 
\delta \rho \right) \equiv
\langle \langle \delta \rho \mid \Omega \mid \delta\rho \rangle \rangle 
\label{a19}
\end{equation}%
provides a function changing monotonically with time. We have namely 
\begin{equation}
\begin{array}{lll}
\partial _{t}M_{\Omega }\left[ \left( \rho -\rho _{0}\right) ,\left( \rho
-\rho _{0}\right) \right] & = & Tr\left( \partial _{t}\delta \rho ^{\dagger
}\Omega \delta \rho \right) +Tr\left( \delta \rho ^{\dagger }\Omega \partial
_{t}\delta \rho \right) \\ 
&  &  \\ 
& = & Tr\left( \delta \rho ^{\dagger }\mathcal{L}^{\dagger }\Omega \delta
\rho \right) +Tr\left( \delta \rho ^{\dagger }\Omega \mathcal{L\delta }\rho
\right) \\ 
&  &  \\ 
& = & Tr\left( \delta \rho ^{\dagger }\Omega (\mathcal{L}^{\ast }\mathcal{%
+L)\delta }\rho \right) \\ 
&  &  \\ 
& = & 2{\rm Re}~Tr\left( \delta \rho ^{\dagger }\Omega \mathcal{L\delta }%
\rho \right) .%
\end{array}
\label{a20}
\end{equation}%
where we have used (\ref{a14}). Using the expansion 
\begin{equation*}
\mid \delta \rho (t)\rangle \rangle =\sum_{\nu }r_{\nu }\exp \left(
\lambda _{\nu }t\right) \mid x_{\nu }\rangle \rangle
\end{equation*}%
we obtain 
\begin{equation}
\partial _{t}M_{\Omega }(\delta \rho ,\delta \rho )=\sum_{\nu }\mid
r_{\nu }\mid ^{2}\left( \lambda _{\nu }^{\ast }+\lambda _{\nu }\right) \exp 
\left[ \left( \lambda _{\nu }^{\ast }+\lambda _{\nu }\right) t\right] \leq 0,
\label{a21}
\end{equation}%
because all eigenvalues have negative real parts. The rate (\ref{a21})
decreases with time and when steady state is reached, the development
ceases. This is thus a proper variational operator which determines the
direction of time towards the eventual steady state. In particular we note
that the pumped system, in general, lacks a concept of thermal equilibrium.

The expression (\ref{a19}) represents the natural extension of the purity to
the case of irreversible time evolution. If we want an entity extensive in
the combination of uncoupled subsystems, we may introduce the entropy-like
expression%
\begin{equation}
S_{\Omega }=\pm \log M_{\Omega };  \label{a21a}
\end{equation}%
the choice of sign depends on the interpretation of the quantity.

\section{Application: pumped two-level system}

We apply the considerations above to the generic driven two-level system
with pumping and decay out of the system. The time evolution equation is now%
\begin{equation}
\partial _{t}\left[ 
\begin{array}{c}
\rho _{22} \\ 
\rho _{21} \\ 
\rho _{12} \\ 
\rho _{11}%
\end{array}%
\right] =\left[ 
\begin{array}{c}
\Lambda _{2} \\ 
0 \\ 
0 \\ 
\Lambda _{1}%
\end{array}%
\right] \\ 
+\left[ 
\begin{array}{cccc}
-\Gamma _{2} & -iV & iV & 0 \\ 
-iV & \left( i\omega -\gamma \right) & 0 & iV \\ 
iV & 0 & -\left( i\omega +\gamma \right) & -iV \\ 
0 & iV & -iV & -\Gamma _{1}%
\end{array}%
\right] \left[ 
\begin{array}{c}
\rho _{22} \\ 
\rho _{21} \\ 
\rho _{12} \\ 
\rho _{11}%
\end{array}%
\right] .%
\label{s1}
\end{equation}%
Here $\omega $ is the detuning of the driving field, $V$ is the Rabi type
coupling between the levels, and $\Gamma $ and $\gamma $ are the population 
and coherence decay rates. These define the time constants $T_{1}$ and $T_{2}$
respectively. The coherence decay satisfies the constraint%
\begin{equation}
\gamma \geq \frac{1}{2}\left( \Gamma _{1}+\Gamma _{2}\right) .  \label{s2}
\end{equation}%
This follows if we set $V=\Lambda =0$ and impose the condition%
\begin{equation}
\rho _{11}(t)\rho _{22}(t)\geq \rho _{12}(t)\rho _{21}(t).  \label{s3}
\end{equation}

\subsection{Analytic results}

It is straightforward to obtain the complex eigenvalues of the matrix in (%
\ref{s1}) and verify that all real parts correspond to decay.
The steady state of (\ref{s1}) is also easily computed, but the ensuing
expression is too unwieldy to allow an intuitive interpretation. One
quantity of interest is the asymptotic population difference%
\begin{equation}
\rho _{22}(\infty )-\rho _{11}(\infty ) = 
\left( \dfrac{\Lambda _{2}}{\Gamma _{2}}-\dfrac{\Lambda _{1}}{\Gamma _{1}}%
\right) \left( 1-\dfrac{\eta ^{2}V^{2}}{\omega ^{2}+\gamma ^{2}+\eta
^{2}V^{2}}\right) .%
\label{s4}
\end{equation}%
The deviations from the pumping difference without coupling is in the form of 
a power-broadened Lorentzian, where the rate of incoherence is measured 
by the dimensionless parameter
\begin{equation}
\eta ^{2}=\frac{2\gamma \left( \Gamma _{1}+\Gamma _{2}\right) }{\Gamma
_{1}\Gamma _{2}}\geq \frac{\left( \Gamma _{1}+\Gamma _{2}\right) ^{2}}{%
\Gamma _{1}\Gamma _{2}}\geq 4.  \label{s5}
\end{equation}

From the result (\ref{s4}) we obtain some general observations:

The population difference goes to zero:

\begin{itemize}
\item If \ $\frac{\Lambda _{2}}{\Gamma _{2}}=\frac{\Lambda _{1}}{\Gamma _{1}}%
.$Then each level loses on the average as much as it gains.

\item If $V\rightarrow \infty .$ In this case the flopping between the
levels is so fast that the population spends roughly half its time on each
level, and consequently their differences are smeared out.
\end{itemize}

On the other hand, the difference goes to the value $\frac{\Lambda _{2}}{%
\Gamma _{2}}-\frac{\Lambda _{1}}{\Gamma _{1}}$ if the coupling becomes
inefficient. This happens when

\begin{itemize}
\item The coupling is weak, $V\rightarrow 0.$

\item The decoherence rate is strong, $\gamma \rightarrow \infty .$
\end{itemize}

\subsection{Numerical results}

The time evolution equation \ (\ref{s1}) contains 7 parameters with
dimensions $\left[ t^{-1}\right] .$ One may be eliminated by scaling the
time rate, but the parameter space is still too large to allow a full
systematic mapping. Some conclusions are, however, readily obtained.

When we pose initial conditions $\rho (0)$ and numerically integrate the 
equations (\ref{s1}) we can see the manifestations of the remarks above. The 
resulting time evolution is exemplified in Fig.1. 
After scaling the physical parametrs in Eq. (47) by a suitably selected time
parameter, we have 6 remaining dimensionless variables. The numerical
values used in the computations are these dimensionless parameters, which
are given in the figures.

\begin{figure}
\center{\includegraphics[width=14cm,height=!]{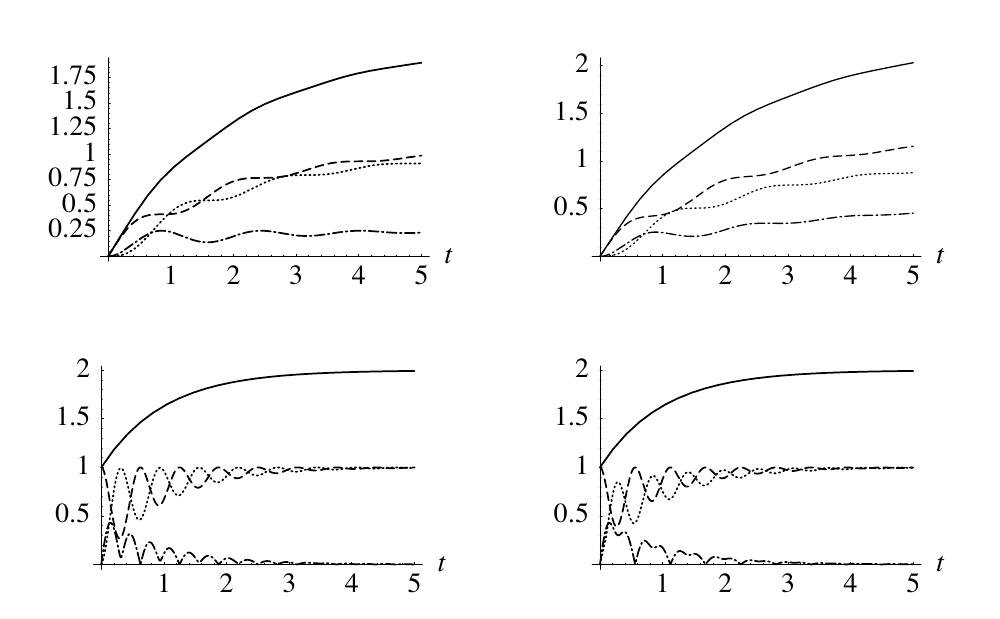}}
\caption {The figures show the total population $\rho_{22}+\rho_{11}$ (solid 
line), $\rho_{22}$ (dashed line), $\rho_{11}$ (dotted line) and the absolue 
value of $\rho_{21}$ (dash-dotted line) as a function of time. In Case 1 (top 
left), the pumping parameters are $\Lambda_1=0$, $\Lambda_2=1.0$, 
the decay rates are $\Gamma_2=0$, $\Gamma_1=1.0$, $\gamma=0.5$, 
the detuning is $\omega=0$ and the coupling $V=2.0$. There is zero 
population in the initial state. In Case 2 (top right) the parameters and initial 
state are the same as in Case 1 but with $\omega=-1.0$. In Case 3 (bottom 
left) the parameters are $\Lambda_1=1.0$, $\Lambda_2=1.0$, 
$\Gamma_2=1.0$, $\Gamma_1=1.0$, $\gamma=1.0$, $\omega=-1.0$ and 
$V=5.0$, and the initial state has $\rho_{22}(0)=1$, with other density matrix 
elements equal to zero. In Case 4 (bottom right) parameters and initial state 
are as in Case 3, but with $\omega=-5.0$. All numerical values are given in 
terms of scaled dimensionless physical variables, as explained in the text.} 
\label{fig1}
\end{figure}

Fast flopping rates $V$ tend to make the final populations equal and the
coherences zero. The decay times are determined by the relaxation rates, 
and during the time evolution the population difference tends to change sign 
$V/\gamma $ times; increasing the detuning, we find that $\omega $ tends to
decrease the amplitude of these oscillations. The effect of $\omega $ is,
however, less than that of the coupling $V.$

In Fig. 2, the functional $M_{\Omega }$ is plotted for the cases in Fig. 1. We 
see that in each case, the expectation value of $\Omega$ decays smoothly 
and monotonically.

\begin{figure}
\center{\includegraphics[width=11cm,height=!]{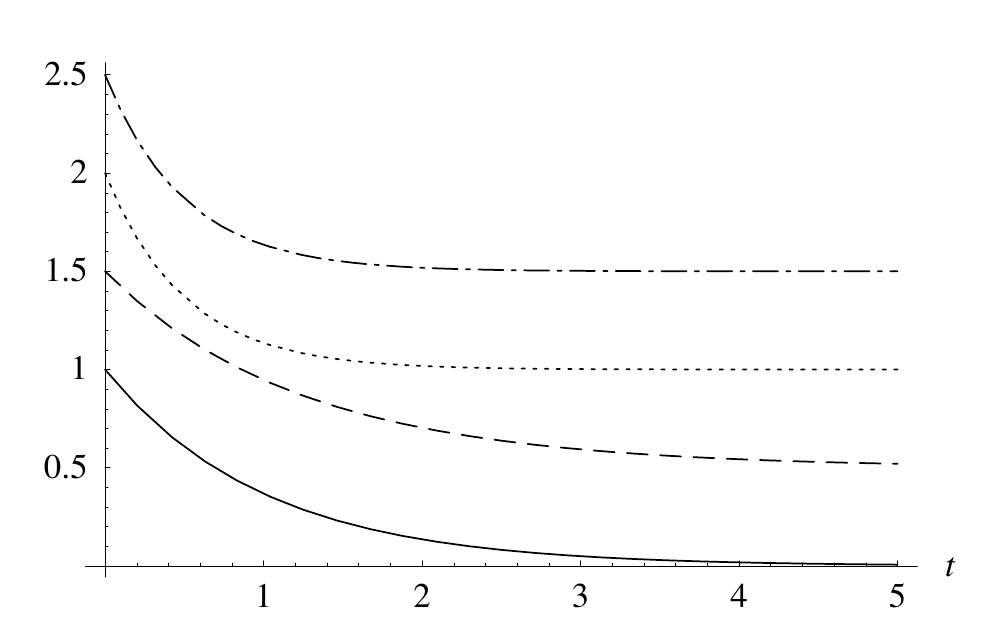}}
\caption {The expectation value of $\Omega$ as a function of time for the 
same four cases as in Fig. 1, from bottom to top Case 1 (solid line), Case 2 
(dashed line), Case 3 (dotted line) and Case 4 (dash-dotted line). In all cases, 
normalisation is chosen as $\langle\rho(0)|\Omega|\rho(0)\rangle=1$. As the 
curves are very similar, they have been shifted vertically by 0.5 with respect 
to each other. The asymptotic value for  the unshifted 
$\langle\rho(t)|\Omega|\rho(t)\rangle$ when 
$t\rightarrow \infty$ is zero in all four cases, and the behaviour is monotonic 
as a function of time. All numerical values are given in terms of scaled 
dimensionless physical variables, as explained in the text.} 
\label{fig2}
\end{figure}

Some details of these four cases chosen as illustration are given in the
Appendix. As seen there, all eigenvalues imply damping as we expect from the
physical interpretation. In all four cases, there are two purely damped
modes. These are the ones describable as pure rate variables.
We have, in fact, integrated a large number of parameter combinations, and
we have found no anomalous behavior; all cases show smooth and regular
behavior. The results are very similar to the ones in the cases presented
here. We expect our analysis to apply to all cases.

\section{Conclusion}

The idea in our line of research has been to find expressions uniquely
indicating the direction of time at each instant. This is particularly
urgent in the field of quantum optics where the time evolution is
conventionally described \ by phenomenological master equations. In
particular, the state space of interest comprises only a part of all
possible states of the physical system. Thus an interpretation in terms of
quantum concepts is natural, but also Markovian rate models fall under our
concepts \cite{Q14}. In fact, the rate equation situation is introducing an
approach \cite{Q33}, which constitutes a precursor to our method.

In mathematics, the related formalism has been known for some time under 
the name of "symmetrization" \cite{Q33a,Q33b}. The Prigogine school
has introduced  a similar description of irreversible time
evolution \cite{Q33c}, but their formalism is based on a different physical 
argument and less concisely defined than our present approach. We base all 
our expressions on precisely defined mathematical concepts. When applicable 
our formulation leads to uniquely determined numerical evaluations.

The approach is limited to linear evolution equations with time-independent
generators. The linearity is natural if we wish time evolution in
independent subsystems to remain uncoupled. Violating this condition takes
us into a totally different category of models. To introduce generators
depending on time is possible, but the ensuing theory then has to deal with
time dependent eigenvalues and eigenelements. This is possible but leads to a
theory too complex to give transparent physical interpretations.

We feel that even with its restrictions, our method is general enough to
illuminate the properties of irreversible system dynamics.

\appendix
\section{Details of the numerical simulations}
Below we give the steady states and the eigenvalues for $\mathcal L$ relating 
to cases 1-4 plotted in Figs. 1 and 2. The steady states $\rho_0$ agree with
Eq. (\ref{s4}), and all real parts of eigenvalues are negative as required. The 
parameters are as follows:
\begin{itemize}
\item {{Case 1:} $\Lambda_1=0$, $\Lambda_2=1.0$, $\Lambda_{21}=0$, 
$\Gamma_2=0$, $\Gamma_1=1.0$, $\gamma=0.5$,  $\omega=0$,  $V=2.0$}\\
$\rho_0 = (1.0625,-0.25i, 0.25i, 1)^T$\\
$\lambda_1 = -0.5-3.9686 i$, $\lambda_2 = -0.5+3.9686 i$, 
$\lambda_3 = \lambda_4 = -0.5$

\item {{Case 2:} $\Lambda_1=0$, $\Lambda_2=1.0$, $\Lambda_{21}=0$, 
$\Gamma_2=0$, $\Gamma_1=1.0$, $\gamma=0.5$,  $\omega=-1.0$,  $V=2.0$}\\
$\rho_0 = (1.3125,-0.5-0.25i, -0.5+0.25i, 1)^T$\\
$\lambda_1 = -0.5000-4.0945 i$, $\lambda_2 = -0.5000+4.0945 i$, 
$\lambda_3 = -0.6221$, $\lambda_4 = -0.3779$

\item {{Case 3:} $\Lambda_1=1.0$, $\Lambda_2=1.0$, $\Lambda_{21}=0$, 
$\Gamma_2=1.0$, $\Gamma_1=1.0$, $\gamma=1.0$,  $\omega=1.0$,  $V=5.0$}\\
$\rho_0 =  (1, 0, 0, 1)^T$\\
$\lambda_1 = -1.0-10.0499i$, $\lambda_2 = -1.0+10.0499 i$, 
$\lambda_3 =  \lambda_4 = -1.0$

\item {{Case 4:} $\Lambda_1=1.0$, $\Lambda_2=1.0$, $\Lambda_{21}=0$, 
$\Gamma_2=1.0$, $\Gamma_1=1.0$, $\gamma=1.0$,  $\omega=5.0$,  $V=5.0$}\\
$\rho_0 = (1,0,0,1)^T$\\
$\lambda_1 = -1.0-11.1803 i$, $\lambda_2 = -1.0+11.1803 i$, 
$\lambda_3 = \lambda_4 = -1.0.$

\end{itemize}

\end{document}